\documentclass[reprint, amsmath, amssymb, aps, superscriptaddress]{revtex4-2}

\usepackage{graphicx}
\usepackage{dcolumn}
\usepackage{bm}
\usepackage{amsmath}
\usepackage{amssymb}
\usepackage[colorlinks,linkcolor=blue,citecolor=blue]{hyperref}

\graphicspath{ {Figs/} }

\begin{document}

\preprint{APS/123-QED}

\title{Emergent Dynamical Ising Transition in Diffusive Sandpiles}
\author{Armin Makani}
\affiliation{Department of Physics, University of Mohaghegh Ardabili, P.O. Box 179, Ardabil, Iran}
\author{M. N. Najafi}
\email{morteza.nattagh@gmail.com}
\affiliation{Department of Physics, University of Mohaghegh Ardabili, P.O. Box 179, Ardabil, Iran}

\date{\today}

\begin{abstract}
Minimally stable site (MSS) clusters play a dominant role in shaping avalanches in the self-organized critical (SOC) systems. The manipulation of MSS clusters through local smoothings (diffusion) alter the MSS landscape, suppressing rare avalanches and postponing them until they manifest as spanning avalanches. By leveraging the Inverse Ising problem, we uncover a duality between diffusive sandpiles and equilibrium statistical physics. Our analysis reveals an emergent magnetic instability in the dual Ising model, coinciding with the formation of spanning avalanches and marking a transition to a correlated percolation regime.  At this point, the MSS loop soups exhibit fractal self-similarity and power-law distributions, while the effective pairwise interactions in the dual system vanish—signaling a magnetic transition characterized by abrupt changes in magnetization and spin susceptibility. Crucially, we show that diffusion fundamentally reshapes avalanche dynamics: the spatial anti-correlations of MSSs in standard SOC systems transform into positive correlations when diffusion is introduced. These findings bridge self-organized criticality, percolation theory, and equilibrium phase transitions, shedding new light on emergent criticality and large-scale correlations in non-equilibrium systems.
\end{abstract}

\maketitle

Sandpiles represent the prototypical example of self-organized criticality (SOC), a phenomenon describing the spontaneous emergence of scale invariance in driven dissipative systems. SOC provides a unifying framework for understanding avalanche dynamics across diverse natural and artificial systems~\cite{dhar1999abelian,najafi2021some}. Despite the apparent simplicity of the underlying rules—and nearly four decades of theoretical and numerical exploration—sandpile models continue to reveal unexpected behaviors and intricate mathematical structures, challenging our understanding of self-organization. The Bak-Tang-Wiesenfeld (BTW) sandpile model~\cite{Bak1987self,Bak1988self}, a paradigmatic example of SOC, exhibits critical behavior through avalanches triggered by the toppling of minimally stable sites (MSS) defined as the sites that become unstable upon a single perturbation~\cite{Bak1988self,Christensen1991dynamical,Bak1994fractals}. The analysis of the corresponding minimally stable states (a stable configuration in which a small perturbation can lead to redistribution of grains) is crucial in real SOC systems like earthquakes~\cite{Bak1995Fractal}, and rainfalls~\cite{Bove2006Complexity}. The spatial organization of these MSS clusters plays a fundamental role in governing the properties of subsequent avalanches, specially the rare events where MSS clusters hypothesized to percolate~\cite{Bak1987self,Wiesenfeld1989sandbox}—a claim that had never been formally proven and is found here to be unsupported. This suggests a key question: can the manipulation of MSS clusters serve as a control mechanism for mitigating or forecasting extreme events in SOC systems?

A particularly effective approach to controlling rare events in SOC involves diffusive sandpiles, where \textit{local smoothing} mechanisms redistribute mass, thereby reducing the likelihood of large avalanches. However, previous studies have shown that such control mechanisms do not eliminate extreme events; instead, they give rise to an intermittent stationary regime marked by recurrent, large-scale spanning avalanches~\cite{najafi2019local}. Given the crucial role of MSS cluster patterns in avalanche formation, an interesting question emerges: how do MSS clusters evolve when local smoothings, as defined in diffusive sandpile models, are introduced? Exploring this question helps illuminate the connection between the internal organization of MSS clusters and the long-term dynamical evolution of the system.

Although SOC systems are inherently out of equilibrium, their dynamics can be mapped onto an equilibrium statistical model via the Inverse Ising problem~\cite{Nguyen2017Inverse,Aurell2012Iverse}, providing an effective thermodynamic framework. The validity of this mapping, requires further investigation by assessing whether the effective Ising model can reproduce the input patterns. By reconstructing both the effective pairwise interactions and the effective magnetic field of an Ising model from sandpile dynamics, we can examine how the system evolves toward the formation of spanning avalanches. In this letter we demonstrate that the onset of spanning avalanches, accompanied by percolative MSS clusters, coincides with a significant shift in the effective Ising interactions and the emergence of magnetic instability in the dual Ising model.\\

The diffusive sandpile model~\cite{najafi2019local} extends the standard sandpile model by incorporating local smoothing, which facilitates the diffusion of energy across the lattice. We consider an $L\times L$ square lattice where each site has a random energy value $1\le \epsilon(i)\le \epsilon_c\equiv 4n$, with  $n$ set to $10$ in this study. At each time step, an energy unit (a single stimulation) is added to a randomly selected site $i$, updating its energy as $\epsilon_i\to \epsilon_i+n$. When the energy of a site exceeds $\epsilon_c$ (becoming unstable), a ``toppling" (local relaxation) occurs through a conservative redistribution of energy: $\epsilon_i\to \epsilon_i+\Delta_{ij}$  where $\Delta_{ij}=-\epsilon_c$ if $i = j$, $\Delta_{ij}=n$ if $i$ and $j$ are neighbors, and zero otherwise. A toppling event at one site can trigger instability in neighboring sites, causing a chain reaction of topplings until the entire lattice reaches a stable configuration. The system operates under open boundary conditions, where energy dissipation occurs at the boundaries. An avalanche is defined as the chain of topplings between two stable configurations. A ``local smoothing" process further enhances stability by selecting a random site and assessing whether its configuration can be stabilized. This is done by examining and redistributing energy based on the energy levels of its neighboring sites. Suppose the chosen site is $i$, with nearest neighbors $\left\lbrace i_k\right\rbrace_{k=1}^4$, and their respective energies are $\left\lbrace \epsilon_{i_k}\right\rbrace_{k=1}^4$. Among these neighbors, let $i_{\text{max}}/i_{\text{min}}$ denote the site with the highest/lowest energy, such that $\epsilon(i_{\text{max}}) = \text{Max}\left\lbrace \epsilon_{i_k}\right\rbrace_{k=1}^4$ and $\epsilon(i_{\text{min}}) =\text{Min}\left\lbrace \epsilon_{i_k}\right\rbrace_{k=1}^4$. A local smoothing operation consists of two updates: An energy transfer of $\delta \epsilon_1\equiv \text{int}\left( [\epsilon(i_{\text{max}})-\epsilon(i)]/2\right) $ units (if positive) from $i_{\text{max}}$ to $i$, and an energy transfer of $\delta \epsilon_2\equiv \text{int}\left( [\epsilon(i)-\epsilon(i_{\text{min}})]/2\right) $ units (if positive) from $i$ to $i_{\text{min}}$, where $\text{int}(x) $ represents the integer part of $x$. If more than one sites have max/min values, the site to transfer from/to is selected randomly. The local smoothing does not occur on unstable sites. The diffusion strength is controlled by a parameter $\zeta$ defined as the ratio $\zeta=\frac{n_2}{n_1}$ where $n_2$ is the number of local smoothings during the time interval of $n_1$ topplings. Defining an MSS as the site with $\epsilon\in [3n+1,4n]$ one can calculate the MSS density as $\rho\equiv\frac{n_{\text{MSS}}}{L^2}$, where $n_{\text{MSS}}$ represents the number of MSS in a system. \\

The diffusive sandpile model for $\zeta>0$ exhibits a time-dependent (intermittent) stationary phase, contrary to the recurrent states of the standard BTW model ($\zeta=0$) where the spatial average energy $\bar{\epsilon}$ is statistically constant. The concept of \textit{spanning avalanches} was coined in~\cite{najafi2019local} for $\zeta>0$, where intermittent stationary states were recovered through which $\bar{\epsilon}$ displays a periodic behavior over time, with each period beginning and ending with a spanning avalanche. A spanning avalanche occur at a threshold where $\bar{\epsilon}$ undergoes a sharp drop, followed by a gradual increase until the next spanning avalanche. In this paper we identify an spanning avalanches by detecting a significant drop in $\bar{\epsilon}$ using a predefined threshold (see Appendix~\ref{sec:sm1} for a statistical analysis of drops). Spanning avalanches should be distinguished from a \textit{percolating avalanche} defined as the avalanche that connects two opposite boundaries. A statistical analysis of these energy drops and their timing was previously conducted in~\cite{najafi2019local}, where a mean-field theory for diffusive sandpiles was developed based on the nonlinear branching ratio. To characterize the MSS statistics we define time ($T$) as the number of energy injections to the system. In the intermittent stationary state spanning avalanches are set as the time origin, i.e. $T$ is set to zero when a spanning avalanche occurs. We analyze the MSS statistics in stable configurations.\\

\begin{figure} 
	\centering
	\includegraphics[scale=0.96]{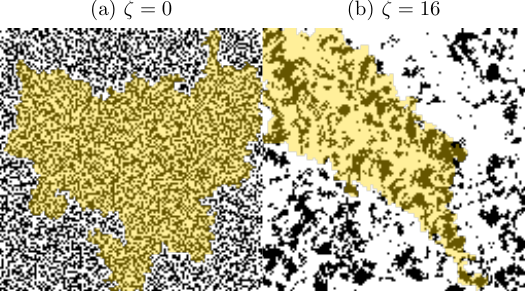}
	\caption{Two MSS configurations for the diffusive sandpiles associated with percolating avalanche with (a)  $\zeta=0$ (Standard BTW model), and (b) $\zeta=16$, The shaded area shows the domain where the subsequent (top to bottom) percolating avalanche has occurred.}
	\label{fig:model_visualization}
\end{figure}
\begin{figure*} 
  \centering 
  \includegraphics[scale=0.66]{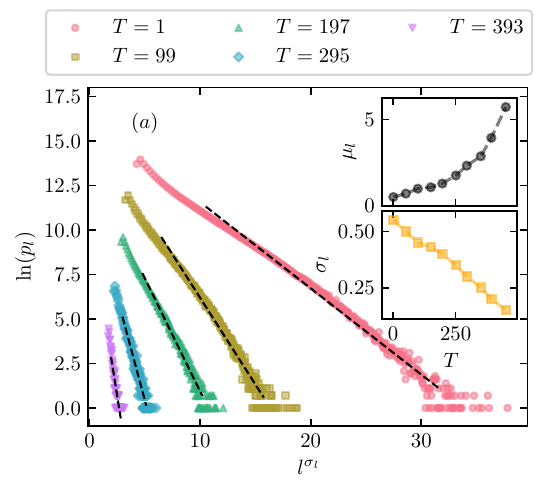}
  \includegraphics[scale=0.66]{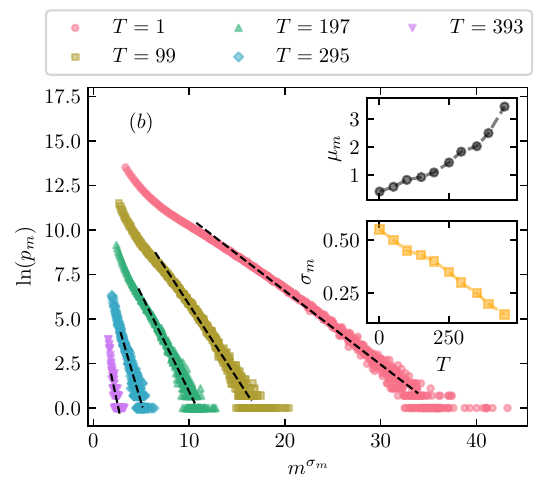}
  \includegraphics[scale=0.63]{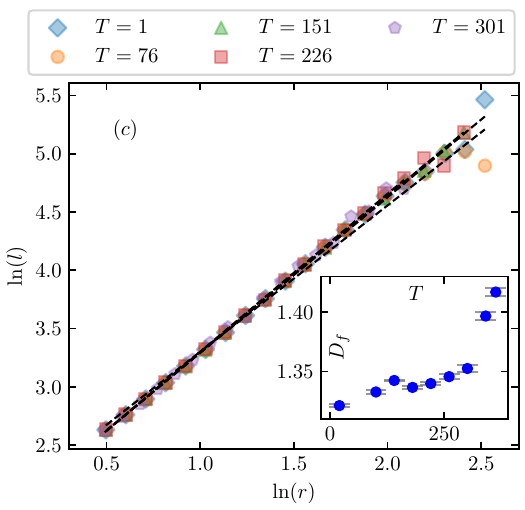}
  \caption{(a) and (b) show the distribution functions $\ln p(x)$ in terms of $\ln x$ for various $T$ values where $x=l$ and $m$ for (a) and (b) respectively, and insets show the exponents $\mu_x$ and $\sigma_x$ defined in Eq.~\ref{Eq:stretched}. (c) shows $\ln l$ in terms of $\ln r$ for various $T$ values, and inset shows the exponent $D_f$ in terms of $T$.}
  \label{fig:LSE}
\end{figure*}

Figure~\ref{fig:model_visualization} illustrates an example of an MSS cluster for $L=128$ with $\zeta=0$ and $\zeta=16$, where black regions represent MSS clusters just before a percolating avalanche. As observed in the figure, activating $\zeta$ introduces spatial correlations in MSS clusters, though their specific configuration depends on $T$. In the standard BTW model ($\zeta=0$) the density remains nearly independent of the corresponding avalanche, with $\rho_{\text{BTW}}^{}=0.440614 \pm 8\times 10^{-06}$. For $\zeta>0$, $\rho$ exhibits a dependence on $T$, which will be analyzed later. To characterize the intermittent stationary regime, we identify MSS clusters using the Hoshen-Kopelman algorithm~\cite{Hoshen1976percolation}, which assigns labels to connected sites on the lattice. This method produces disconnected clusters, whose boundaries form loops. The collection of independent loops is referred to as the \textit{loop soup ensemble}, over which we compute statistical properties. For each connected cluster, the external perimeter is defined as the loop length $l$, while its gyration radius is given by $r\equiv \sqrt{\frac{1}{l}\sum_{i=1}^l \left| \textbf{r}_i-\textbf{r}_{\text{com}}\right|^2}$, where the center of mass is $\textbf{r}_{\text{com}}\equiv \frac{1}{l}\sum_{i=1}^l  \textbf{r}_i$. The corresponding mass $m$ is defined as the number of black sites enclosed within the loop. We analyze the distribution functions of these quantities for the loop soup ensemble in Fig.~\ref{fig:LSE}. Figures~\ref{fig:LSE}a and~\ref{fig:LSE}b show the probability distributions $p(x)$, $x=l,m$ respectively. Our analysis indicates that these distributions are best described by \textit{stretched exponentials}, defined as
\begin{equation}
p(x)\propto \exp\left[-\mu_x x^{\sigma_x}\right],
\label{Eq:stretched}
\end{equation}
where $\mu_x$ and $\sigma_x$ are the corresponding exponents for $x=l,m$, which generally depend on $T$. We observe that $\mu_x$ ($\sigma_x$) increases (decreases) monotonically with $T$. Notably, the distribution functions approach a power-law form as $(\mu_x^{-1},\sigma_x)\to 0$ leading to the scaling behavior $\left. p(x)\right|^{}_{(\mu_x^{-1},\sigma_x)\to 0}\propto x^{-\mu_x\sigma_x}$. The insets of Figs.~\ref{fig:LSE}a and~\ref{fig:LSE}b confirm that this condition is met for sufficiently large $T$, indicating that the distribution functions transition to a power-law form over long timescales. Stretched exponential functions belong to the class of fat-tailed distributions when $\sigma_x<1$, and appear in a variety of systems, including the relaxation time distribution in electronic glasses~\cite{Phillips1996stretched}, the size distribution of avalanches in active piles~\cite{Najafi2023anomalous}, and polymer dynamics~\cite{Cherayil1992Stretched}. These distributions are often considered a hallmark of memory and correlation effects~\cite{Sun2018Investigation}, suggesting that such effects become significant for $\zeta>0$.
The correlation between 
$l$ and $r$ follows a scaling relation (Fig.~\ref{fig:LSE}c) given by $l\propto r^{D_f}$, where $D_f$ is another $T$-dependent scaling exponent, known as the \textit{fractal dimension} in scale-invariant systems (inset). For small $T$, this exponent is consistent with the fractional value $\frac{4}{3}$, while for sufficiently large times, it converges to a value close to $\frac{11}{8}$. Notably, the fractal dimension of the external perimeter of Fortuin–Kasteleyn clusters in the two-dimensional critical Ising model is also $\frac{11}{8}$~\cite{janke2004geometrical}, suggesting a possible connection between the two systems.

\begin{figure} 
	\centering
	\includegraphics[scale=0.75]{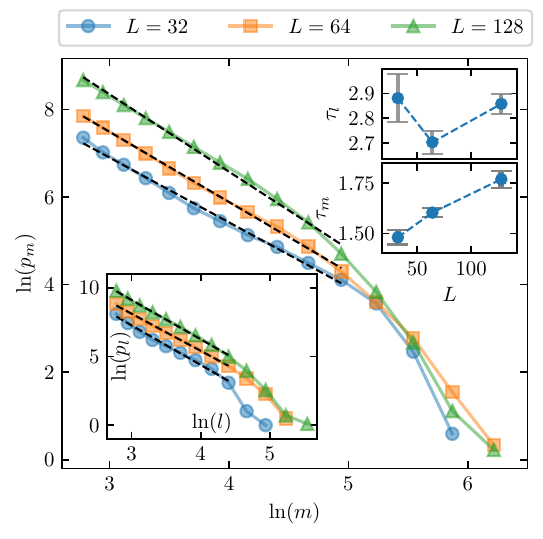}
	\caption{The distribution function of percolative MSS cluster mass $m$ (main), and loop length $l$ (lower inset). The upmost and the middle insets show the corresponding exponents $\tau_l$ and $\tau_m$ in terms of $L$ respectively.}
	\label{fig:LoopMassDist}
\end{figure}

These arguments suggest that the MSS loop soups become self-similar when percolative avalanches appear (in sufficiently large times), during which the system undergoes global relaxation accompanied by the emergence of spanning avalanches. To examine this idea more directly, we analyze the MSS loop soups just before percolative avalanches—specifically, the MSS configurations one step prior to such an event—which we refer to as \textit{percolative MSS loop soups}. The results presented in Fig.~\ref{fig:LoopMassDist} indicate that these percolative MSS loop soups exhibit self-similarity, displaying power-law behavior in various observables, such as $l$ and $m$ with scaling exponents determined to be (insets) $\tau_l=2.85\pm 0.04$ and $\tau_m=1.75\pm 0.04$ for $L=128$. This observation is consistent with the above arguments, suggesting that the percolative MSS loop soup is self-similar.
This observation raises a crucial question: How does the system approaches the percolative regime over time? To gain a better insight, one may search for a connection to an equilibrium system with local binary states, distinguishing between minimally stable and non-minimally stable states. The \textit{inverse Ising problem} is a tool to map our binary state system to an Ising model, with tuned coupling constants between auxiliary spins. Despite the fundamental differences between out-of-equilibrium and equilibrium systems, we establish a direct mapping between our system and the Ising model by interpreting MSS's as ``spin-up" states, while the remaining sites correspond to ``spin-down" ones. While such mappings are generally nontrivial due to the lack of detailed balance in the sandpile dynamics~\cite{Nagel1992Instabilities}, our approach successfully reconstructs an effective Ising-like representation.\\

The Ising energy of a ``spin" configuration $\left\lbrace s_i\right\rbrace_{i=1}^N$ is given by
\begin{equation}
	\mathcal{H}_{\textbf{J},\textbf{h}}(\boldsymbol{s}) =- \sum_{ij} J_{ij} s_i s_j - \sum_{i} h_i s_i,
\label{Eq:Hamiltonian}
\end{equation}
where $s_i=+1$ and $-1$ for spin up and spin down sites respectively, $\textbf{J}\equiv \left\lbrace J_{ij}\right\rbrace_{i,j=1}^N$ is the set of pairwise coupling constant, and $\textbf{h}\equiv \left\lbrace   h_i\right\rbrace_{i=1}^N$ is the magnetic field. The temperature is implicitly incorporated into the coupling constants \( J_{ij} \) and the external fields \( h_i \), thereby characterizing the thermal fluctuations in the system. The connection of sandpiles to the Ising model is implemented by assigning $s=+1$ to an MSS, and $s=-1$ to the rest. The magnetization is defined as 
\begin{equation}
\bar{s}\equiv L^{-2}\sum_is_i=2\rho-1,
\end{equation}
which results to $\bar{s}(\zeta=0)\equiv\bar{s}_{\text{BTW}}^{}=-0.11878\pm 0.16\times 10^{-5}$. Unlike the direct Ising problem, where magnetization and spin correlations are computed for given external fields and interactions, the inverse Ising problem involves inferring the underlying coupling constants and magnetic field from a given set of spin configurations. Putting in another way, for a given external magnetization and pairwise correlations—represented here by MSS configurations—one aims to determine the corresponding Ising model parameters, and therefore poses a significant challenges, see~\cite{Nguyen2017Inverse} for details, where various approaches, primarily grounded in the maximum likelihood framework, have been introduced. \\

The equilibrium properties of the Ising model are governed by the Boltzmann distribution
\begin{equation}
\mathcal{P}_{\textbf{J},\textbf{h}}(\textbf{s}) = \frac{\exp \left(-\mathcal{H}_{\textbf{J},\textbf{h}}(\textbf{s})\right)}{\mathcal{Z}_{\textbf{J},\textbf{h}}},
\label{Eq:probMeas}
\end{equation}
where $\mathcal{Z}_{\textbf{J},\textbf{h}}=\exp\left[-\mathcal{F}_{\textbf{J},\textbf{h}}\right]$ is the partition function, and $\mathcal{F}_{\textbf{J},\textbf{h}}$ is the free energy of the system. Using this probability measure, the average of any statistical observable can be expressed as $\left\langle O(\textbf{s})\right\rangle\equiv \sum_\textbf{s}\mathcal{P}_{\textbf{J},\textbf{h}}(\textbf{s})O(\textbf{s})$. In particular, the magnetization and pairwise correlations in the Ising model are given by $m_i^{(\textbf{J},\textbf{h})}\equiv \left\langle s_i\right\rangle= \sum_\textbf{s}\mathcal{P}_{\textbf{J},\textbf{h}}(\textbf{s})s_i$, and $\chi^{(\textbf{J},\textbf{h})}_{ij}\equiv \left\langle s_is_j\right\rangle= \sum_\textbf{s}\mathcal{P}_{\textbf{J},\textbf{h}}(\textbf{s})s_is_j$. Alternatively, the empirical averages can be computed from a given set of $M$ independent spin samples $\mathcal{S}\equiv \left\lbrace\textbf{s}^{\mu} \right\rbrace $, where $\mu=1,2,...,M$, as $m_i^{\mathcal{S}}\equiv \left\langle s_i\right\rangle^{\mathcal{S}}= M^{-1}\sum_\mu s_i^\mu$, and $\chi^{\mathcal{S}}_{ij}\equiv \left\langle s_is_j\right\rangle^{\mathcal{S}}= M^{-1}\sum_\mu s_i^{\mu}s_j^{\mu}$. The effectiveness of an inverse Ising model is determined by how well these empirical averages match their theoretical counterparts. \\

The \textit{equilibrium inverse Ising problem} addresses two key questions: (1) Do the given samples follow the equilibrium (Boltzmann) probability measure in Eq.~\ref{Eq:probMeas}? (2) If so, what set of parameters $(\textbf{J},\textbf{h})$ best describe this measure? Both questions present significant challenges. For the first, a common approach is to assume that the duality holds. One then determines the optimal parameters, denoted as $\textbf{J}^*$ and $\textbf{h}^*$, and subsequently checks whether the resulting statistical observables exhibit the expected patterns. We successfully captured the essential statistical correlations of the original system, i.e. re-built the configurations from the diffusive sandpiles using the effective Ising model. We also find that the onset of percolative regime aligns with magnetic instabilities in the Ising dual, reinforcing the validity of our mapping. The second question is more technical and will be addressed in the following discussion. The log-likelihood of the model parameters, given the observed configurations $\mathcal{S}$, is defined as follows: 
\begin{equation}
\mathcal{L}_{\mathcal{S}}\left(\textbf{J},\textbf{h}\right)\equiv \sum_{i<j}J_{ij}\chi^{(\mathcal{S})}_{ij}+\sum_ih_im^{(\mathcal{S})}_{i}- \mathcal{F}_{\textbf{J},\textbf{h}},
\end{equation}
i.e. it is a Legendre transformation of the free energy with respect to both $\textbf{J}$ and $\textbf{h}$. A most direct method to find $\textbf{J}^*$ and $\textbf{h}^*$ is to maximize $\mathcal{L}_{\mathcal{S}}\left(\textbf{J},\textbf{h}\right)$ with respect to $\textbf{J}$ and $\textbf{h}$: $\left\lbrace \textbf{J}^*,\textbf{h}^*\right\rbrace=\text{max}_{\textbf{J},\textbf{h}}\mathcal{L}_{\mathcal{S}}\left(\textbf{J},\textbf{h}\right) $. Noting that
\begin{equation}
\begin{split}
\frac{\partial \mathcal{L}_{\mathcal{S}}}{\partial h_i}=m^{(\mathcal{S})}_{i}-m_i^{(\textbf{J},\textbf{h})},\ \frac{\partial \mathcal{L}_{\mathcal{S}}}{\partial J_{ij}}=\chi^{(\mathcal{S})}_{ij}-\chi_{ij}^{(\textbf{J},\textbf{h})}.
\end{split}
\end{equation}
one finds the maximal $\mathcal{L}_{\mathcal{S}}$ corresponds to the conditions $m_i^{(\mathcal{S})}=m_i^{(\textbf{J},\textbf{h})}$, and $\chi^{(\mathcal{S})}_{ij}=\chi_{ij}^{(\textbf{J},\textbf{h})}$ as pointed out before. A gradient-descent algorithm called Boltzmann machine learning (ML) is based on these equations~\cite{Nguyen2017Inverse}.\\
In this paper, we employ nodewise \( l_1 \)-regularized logistic regression as introduced by Ravikumar et al.~\cite{Ravikumar2010High}, for the case of zero magnetic field. In this approach, each spin $s_g$ is treated as the dependent variable, while all other spins serve as predictors. For further details, see Appendix\ref{sec:sm2}. A key challenge of this direct method is its computational complexity, making it prohibitively expensive to apply to large lattice sizes. To address this, we also utilize the Sessak-Monasson (SM) approximation~\cite{Sessak2009Small}, which relies on a perturbative expansion of $\mathcal{F}_{\textbf{J},\textbf{h}}$ in terms of the correlation functions \( c_{ij} = \chi^{(\mathcal{S})}_{ij} - m^{(\mathcal{S})}_i m^{(\mathcal{S})}_j \). This method involves iterative Legendre transformations of the Helmholtz free energy with respect to both the couplings and external fields, ultimately yielding the free energy under fixed magnetization and correlation conditions. Within the SM framework, the pairwise interactions $J^*_{ij}$ and the effective fields $h^*_i$ are expressed in terms of \( L_i = 1 - \left(m_i^{\mathcal{S}}\right)^2, \quad K_{ij} = \frac{c_{ij}}{L_i L_j} \) to up to an arbitrary order of $K$. In our analysis, we truncate the expansion at 
O($K^4$)~\cite{Sessak2009Small}. For further details, see Appendix~\ref{sec:sm2}.\\

\begin{figure} 
  \centering
\includegraphics[width=8cm, height=8cm]{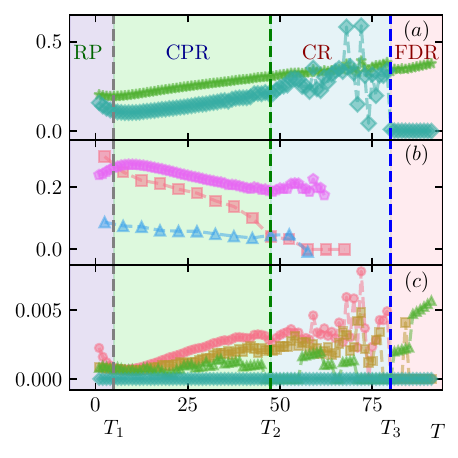}
  \caption{The MSS phase space of diffusive sandpiles in terms of $T$ for $L=128$ and $\zeta=16$. The abbreviates refer to: RP $\to$ refractory period, CPR $\to$ correlated percolation regime, CR $\to$ crossover regime, and FDR $\to$ field dominated regime. (a) shows density $\rho$ ($\star$) and susceptibility ($\diamond$), obtained directly from the diffusive sandpile samples, (b) shows the parameters of the inverse Ising problem: $-0.6\times \bar{h}$ ($\triangle$), $J_{\text{NN}}^{\text{SM}}$ ($\square$) and $J_{\text{NN}}^{\text{ML}}$ ($\triangledown$), and (c) shows the percolation probability for various $L$ values ($L=16,32,64$ and $128$ from top to bottom).}
  \label{fig:phase_space1}
\end{figure}

Figure~\ref{fig:phase_space1} represents the phase diagram of MSS clusters in terms of time $T$. Figure~\ref{fig:phase_space1}a shows $\rho^{(\mathcal{S})}\equiv \frac{1}{2}\left(\bar{m}^{(\mathcal{S})}+1\right)$ and susceptibility $\chi^{(\mathcal{S})}\equiv \left\langle \bar{s}^2 \right\rangle- \left(\bar{m}^{\mathcal{S}}\right)^2$ in terms of $T$, where $\bar{m}^{(\mathcal{S})}\equiv \left\langle \bar{s}\right\rangle=L^{-2}\sum_im^{(\mathcal{S})} $.  Figure~\ref{fig:phase_space1}b illustrates the results obtained using the inverse Ising approach: the nearest neighbor pairwise coupling constant obtained using ML ($J_{\text{NN}}^{\text{ML}}$), and SM ($J_{\text{NN}}^{\text{SM}}$) methods and also the average magnetic field obtained using the SM method $\bar{h}^{\text{SM}}\equiv L^{-2}\sum_{i=1}^{L^2}h_i^{\text{SM}}$ are reported in this figure (note that $-0.6\bar{h}^{\text{SM}}$ is represented to be compatible with the other plots). The percolation probability in Fig.~\ref{fig:phase_space1}c is associated with percolative MSS clusters normalized by the total number of clusters. Four regions are distinguishable (all numbers are for $L=128$):
\begin{itemize}
    \item \textbf{Refractory Period} $0\le T\lesssim T_1$ ($T_1\in [6-10]$): During this interval $\rho^{(\mathcal{S})}$ and $\chi^{(\mathcal{S})}$ decrease (Fig.~\ref{fig:phase_space1}a), and $\bar{h}^{\text{SM}}$ decreases towards more negative values, favoring negative spins, corresponding to non-MSS's (Fig.~\ref{fig:phase_space1}b). Correspondingly the percolation probability decreases in this region. The word ``refractory" refers to the fact that the system resists an increase in the number of MSS's while the average height grows due to local smoothings.
    \item \textbf{Correlated Percolation Region} $T_1\lesssim T\lesssim T_2$ ($T_2\in [45-50]$): $\rho^{(\mathcal{S})}$ and $\chi^{(\mathcal{S})}$ (Fig.~\ref{fig:phase_space1}a), $\bar{h}^{\text{SM}}$ (Fig.~\ref{fig:phase_space1}b), and percolation probability (Fig.~\ref{fig:phase_space1}c)  increase almost linearly with time, while $J_{\text{NN}}^{\text{ML}}$ and $J_{\text{NN}}^{\text{SM}}$ (Fig.~\ref{fig:phase_space1}b) monotonically decreases. The word ``Correlated" refers to the fact that $J_{\text{NN}}$ is large in this interval. 
    \item \textbf{Crossover Region} $T_2\lesssim T\lesssim T_3$ ($T_3\in [80-90]$): $\rho^{(\mathcal{S})}$ and $\chi^{(\mathcal{S})}$ and $\bar{h}^{\text{SM}}$, and percolation probability exhibit instability, characterized by large fluctuations (The fluctuations of $\bar{h}^{\text{SM}}$ and $J_{\text{NN}}^{\text{SM}}$ are not shown here to control the plot domain).
    \item \textbf{Field-Dominated Region} $T\gtrsim T_3$: The $\chi^{\mathcal{S}}$ and $J_{\text{NN}}^{\text{SM}}$ and $J_{\text{NN}}^{\text{ML}}$ drop to zero, while non-zero $\bar{h}$ dominates breaking $\mathbb{Z}_2$ symmetry, and making bias (leading to negative magnetization). The percolative and spanning avalanches often take place in this region.
\end{itemize}
Note that the decrease of 
$J_{\text{NN}}$ and the increase of $\bar{h}$ in the correlated percolation region corresponds to a reduction in the number of majority (negative) spins, which represent non-MSSs, leading to an increase in the magnetization. Moreover, $J_{\text{NN}}^{\zeta=0}$ is found to be $-0.1$ for the BTW model which shows that the dual Ising model is in the antiferromagnetic regime. \\

To gain deeper insight into the regimes observed in Fig.~\ref{fig:phase_space1}, we analyze the system at the mean-field level. The number of MSS's, $n_{\text{MSS}}$ is governed by two competing dynamics:
\begin{itemize}
    \item \textbf{Growth Mechanism}: The rate of growth of $n_{\text{MSS}}$ depends on two energy incomes: the external slow drive, and the energy units coming from the neighbors. The rate associated with the external drive is a constant $\alpha$ (related to the rate of the external input), while the second source is related to the average height which is linear with time~\cite{najafi2019local}, i.e. $\beta T$.
    \item \textbf{Decay Mechanism}: $n_{\text{MSS}}$ decays exponentially with a rate proportional to $\Pi(n_{\text{MSS}})$ the probability that a site belongs to an MSS cluster. This occurs because local smoothings remove energy from an MSS, converting it into a non-MSS. The probability $\Pi(n_{\text{MSS}})$ scales linearly with $n_{\text{MSS}}$, following $\Pi(n_{\text{MSS}})=\gamma n_{\text{MSS}}$, where $\gamma$ is a proportionality constant.
\end{itemize}
These dynamics are reflected in the following linear equation:
\begin{equation}
\frac{\text{d}n_{\text{MSS}}}{\text{d}T}=\alpha+\beta T-\gamma n_{\text{MSS}}. 
\label{Eq:MF}
\end{equation}
The solution of this equation
\begin{equation}
n_{\text{MSS}}=\frac{\alpha\gamma-\beta}{\gamma^2}+\frac{\beta}{\gamma} T+C e^{-\gamma T}
\label{Eq:Solusion}
\end{equation}
(where $C$ is a constant) divides the dynamics into two parts: $\frac{\text{d}n_{\text{MSS}}}{\text{d}T}<0$ which corresponds to the refractory period, and $\frac{\text{d}n_{\text{MSS}}}{\text{d}T}>0$ and $n_{\text{MSS}}\ll L^2$ which is the characteristics of the correlated percolation regime. Equation~\ref{Eq:Solusion} predicts a minimum at $T=T_m\to T_1$, which is given by $e^{-\gamma T_m}=\frac{\beta}{C\gamma^2}$, resulting to a minimum value $n_{\text{MSS}}^{(m)}=\frac{1}{\gamma}\left(\alpha+\beta T_m\right)$. For large enough times (small enough to meet the condition $n_{\text{MSS}}\ll L^2$ ), the effect of exponentially decaying term (the main ingredient of the refractory period) disappears, and $n_{\text{MSS}}$ increases almost linearly up to a point where percolating and spanning avalanches form $n_{\text{MSS}}$ becomes comparable to $ L^2$. This is precisely what we see from Fig.~\ref{fig:phase_space1}a. The linear response theory Eq.~\ref{Eq:MF} is valid only up to this point, beyond which the non-linear response of the system play role, where spanning avalanches take place. 

\begin{figure} 
  \centering
  \includegraphics[scale=0.8]{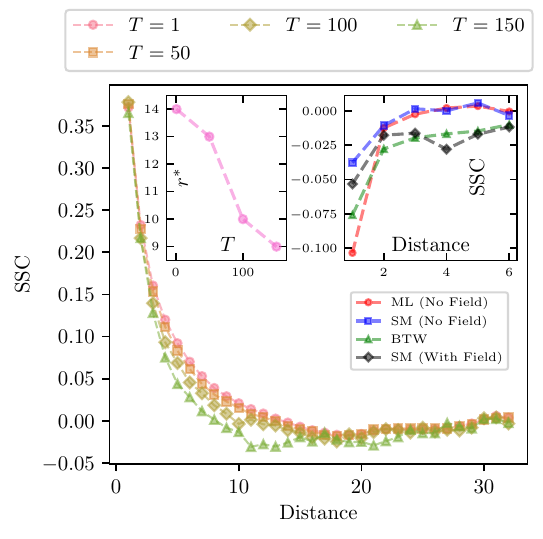}
  \caption{The spin-spin correlation as a function of distance. The main shows diffusive sandpile with $\zeta=16$ for various $T$ values, and the right inset shows the BTW sandpile (the results of SM methd, ML methd and directly using the BTW configurations). The left inset shows the change-of-sign point $r^*$ in terms of $T$.}
  \label{fig:SSC}
\end{figure}

The spin-spin correlation reveals a key distinction between the BTW model, and cases with nonzero $\zeta$, as shown in Fig.~\ref{fig:SSC}. In the BTW model (the inset), the correlation remains negative at all distances, whereas for $\zeta>0$ (main) spins exhibit positive correlation and change sign at a characteristic distance $r^*$. This indicates a fundamental difference in MSS configurations between the two cases: The spatial correlations between ready-to-trigger agents is negative in non-diffusive system (ordinary BTW model), i.e. indicating that these agents decrease the activity of their neighbors, while in a diffusive SOC system, there are positive correlations between these agents leading to a percolative behavior. The left inset shows that $r^*$ decreases monotonically with $T$, suggesting that the positive correlation effects grow over time. \\

To conclude, consider a SOC system exhibiting scale-free avalanche dynamics, and suppose diffusion is introduced to regulate/control rare events. In this paper, we uncovered a deep connection between this problem, percolation theory, and equilibrium statistical mechanics, and tracked the system evolution through mapping it to an Ising model using Inverse Ising problem techniques. Our results reveal a structural transition upon introducing diffusion: The anti-correlation between ready-to-trigger sites (MSS) turns to a positive correlation when system becomes diffusive. Over time, these correlations grow, driving the system toward an unstable state in which the dual Ising model exhibits unpredictable fluctuations in susceptibility and magnetic field. This state corresponds to the onset of large percolative avalanches, where MSS clusters right before a percolative avalanche become self-similar, characterized by the scaling exponents reported in this paper. Eventually, these correlations culminate in spanning avalanches, leading to a global relaxation of energy and a significant drop in the system’s average energy.

Our model not only bridges the gap between SOC’s nonequilibrium dynamics, equilibrium statistical mechanics, and percolation theory but also reveals a secondary fractal structure and emergent long-range correlations in MSS clusters. These findings deepen the understanding of SOC systems and open new avenues for exploring emergent statistical physics in complex systems.

\appendix

\section{Statistical properties of Diffusive Sandpile Model}\label{sec:sm1}

\begin{figure}[htbp]
    \centering
    \includegraphics[width=0.35\textwidth]{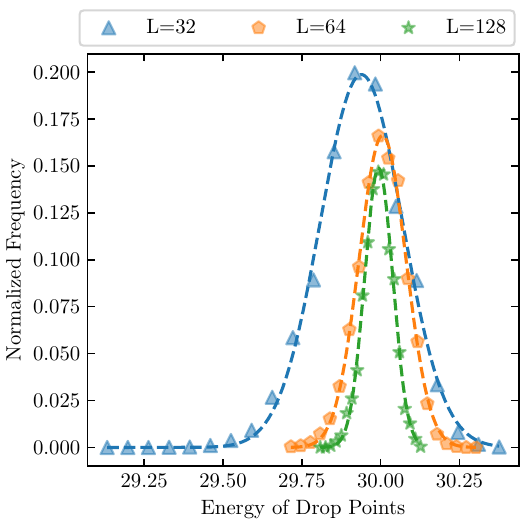}
    \caption{The distribution of drops for lattice sizes of 32, 64, and 128 in the DSM with \(\zeta \neq 0\).}
    \label{fig:sm2}
\end{figure}

\begin{figure}[htbp]
    \centering
    \includegraphics[width=0.35\textwidth]{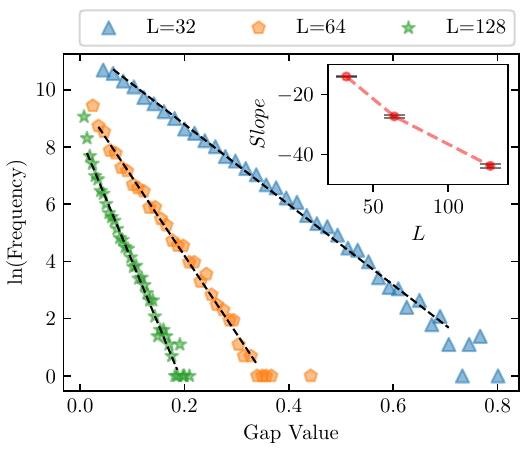}
    \caption{The distribution of gap values for lattice sizes of 32, 64, and 128 in the DSM with \(\zeta \neq 0\).}
    \label{fig:sm3}
\end{figure}

This section is devoted to presenting some aspects of the diffusive sandpile model (DSM). This includes the statistics of energy drops in the stationary regime, and the statistics of the lifetime of avalanches between two successive drops. \\
\begin{figure}[htbp]
    \centering
    \includegraphics[width=0.35\textwidth]{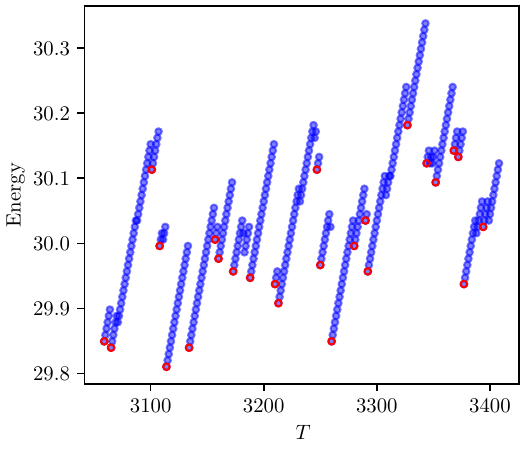}
    \caption{Energy drops (indicated by the red markers) occurring in the DSM for \(\zeta \neq 0\).}
    \label{fig:sm1}
\end{figure}
Fig. \ref{fig:sm1} provides a zoomed-in view of the time series for the $\zeta=16$ DSM within the time interval $[3000-3400]$, highlighting the energy drops characteristic of the intermittent stationary state. Each energy drop is accompanied by a spanning avalanche, marked by red points in the figure. The distribution functions of the energy at the drop points and the energy gaps are presented in Fig. \ref{fig:sm2} and Fig. \ref{fig:sm3}, respectively. The distribution function of the energy at the drop points is Gaussian around an average, while the distribution of the gaps decays exponentially with an $L$-dependent exponent (inset). 

\begin{figure}[htbp]
    \centering
    \includegraphics[width=0.35\textwidth]{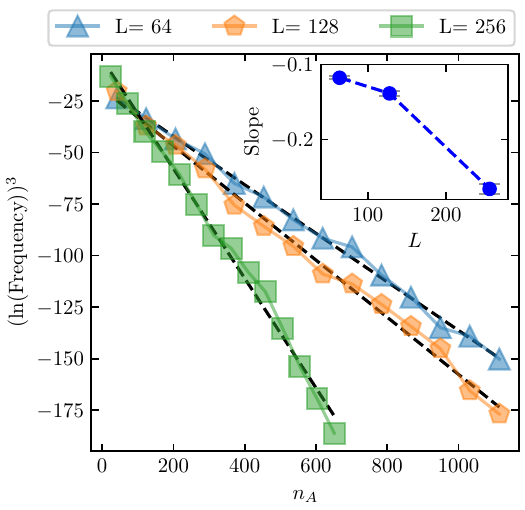}
    \caption{The distribution of the number of avalanches occurring in each interval \(n_A\) for various lattice sizes \(L=64,128\) and \(256\) and \(\zeta = 16\). The slopes of the regression lines are displayed in the inset.}
    \label{fig:sm4}
\end{figure}

\begin{figure}[htbp]
    \centering
    \includegraphics[width=0.35\textwidth]{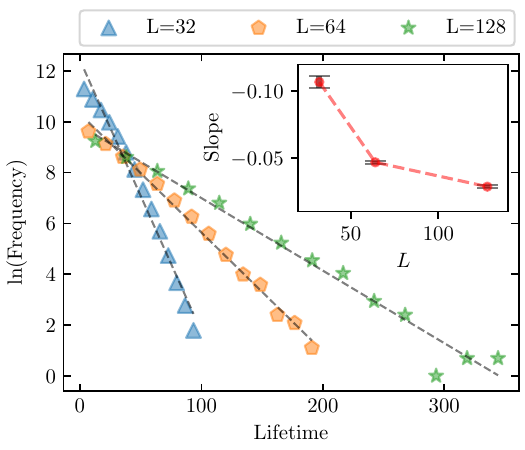}
    \caption{The distribution plot of lifetimes for various lattice sizes in the DSM with \(\zeta = 16\).}
    \label{fig:sm5}
\end{figure}

The number of avalanches in a single interval, indicated by \( n_A \), is analyzed in Fig. \ref{fig:sm4}. The distribution function $P(n_A)$ shows a different behavior: \textit{stretched exponential} form is observed
\begin{equation}
P(n_A)\propto\exp\left[-\alpha_{n_A}n_A^{\frac{1}{3}}\right],
\end{equation}
where $\alpha_{n_A}$ is an exponent associated with decay rate, and the other exponent $\frac{1}{3}$ shows that the distribution function is fat-tailed. The inset of Fig. \ref{fig:sm4} shows the slope ($-\alpha_{n_A}$) in terms of the system size. The diSCtribution function of lifetimes (the duration of each time interval) is represented in Fig. \ref{fig:sm5}, where an exponential decay is observed.\\

\begin{figure}[htbp]
    \centering
    \includegraphics[width=0.35\textwidth]{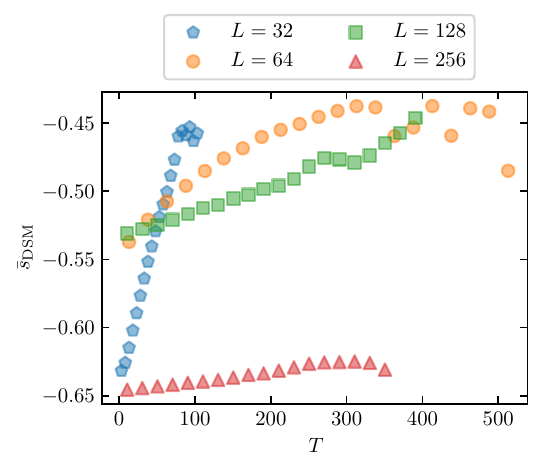}
    \caption{The magnetizations of various DSM models with \(\zeta = 16\) are shown over time for different lattice sizes.}
    \label{fig:sm7}
\end{figure}

\begin{figure}[htbp]
    \centering
    \includegraphics[width=0.35\textwidth]{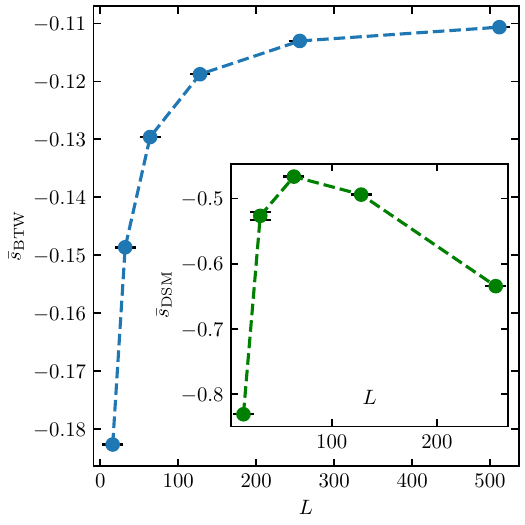}
    \caption{The magnetizations for various DSM models (\(\zeta = 0\)) are shown for different lattice sizes in the main plot, with the associated standard errors indicated by black bars. In the inset plot, the mean magnetizations over time for DSM models with \(\zeta = 16\) are displayed for different lattice sizes, with their corresponding standard errors also represented by black bars.}
    \label{fig:sm6}
\end{figure}

We map the configuration of MSS's to a magnetic system, using the "spin" variable defined in the argument following Eq. \ref{Eq:Hamiltonian} in the main text, i.e. MSS’s are interpreted as “spin-up” ($s=+1$) states, while the remaining sites correspond to “spin-down” ($s=-1$) ones. Figure \ref{fig:sm7} represents the magnetization for $\zeta=16$ for various $L$ values as a function of $T$. We observe that the average magnetization grows with $T$, showing that the number of MSS's increases as the average height of DSM increases.
Figure \ref{fig:sm6} shows the behavior of the magnetization averaged over a time interval for two cases: \(\zeta = 0\) and \(\zeta = 16\). To account for the influence of time in the \(\zeta = 16\) case, the magnetization was calculated at each time, and the results were subsequently averaged over all time points. Note that the time intervals are not defined for $\zeta=0$. As depicted, the magnetization for the \(\zeta = 0\) state converges to approximately -0.11. However, such convergent behavior is not observed in the \(\zeta = 16\) state. 
\section{The Inverse Ising Problem}\label{sec:sm2}
We consider the Ising model consisting of \( N \) spins on top of a graph $G(V,E)$, where $V$ is the set of all nodes in the system, so that $N=|V|$, and $E\equiv \left\lbrace e_{ij}\right\rbrace_{i,j=1}^N $ is the set of all edges. The Hamiltonian is expressed as
\begin{equation}
H(\boldsymbol{s}) = -\sum_{\left\langle ij\right\rangle } J_{ij} s_i s_j - \sum_{i} h_i s_i,
\label{eq:sm2}
\end{equation}
where $\left\langle ij\right\rangle $ shows $i$ and $j$ are connected ($e_{ij}=1$),  \( \boldsymbol{s} = [s_i \in \{-1, 1\} \ \text{for} \ 1 \leq i \leq N] \) represents the set of spin variables, \( \mathbf{J} \equiv \left\lbrace J_{ij}\right\rbrace_{i,j=1}^N \) is the vector of pairwise interaction coefficients, and \( \mathbf{h} \equiv \left\lbrace h_i\right\rbrace_{i=1}^N \) denotes the vector of external magnetic fields. The temperature \( T \) is implicitly incorporated into the coupling constants \( J_{ij} \) and the external fields \( h_i \), thereby characterizing the thermal fluctuations in the system.

The direct (or forward) Ising problem involves calculating the system’s Hamiltonian when the fields and interactions are specified. In contrast, the inverse Ising problem seeks to infer the underlying interaction parameters and external fields from observables, such as magnetization and pairwise correlations, obtained from statistically independent spin configurations. Various approaches, primarily grounded in the maximum likelihood framework, have been developed to address this complexity. Here we describe two methods: the Sessak-Monasson (SM) approximation and the Nodewise \( \ell_1 \)-regularized logistic regression, to solve the inverse Ising problem. 

\subsection{SM Method}\label{sec:sm3}

Consider a set of \( M \) independent spin configurations \( \mathcal{S} \equiv \{ \mathbf{s}^{\mu} \}_{\mu=1}^{M} \). The SM approach applies a perturbative expansion of the Ising model's free energy in terms of the connected correlation functions \( c_{ij} = \chi_{ij}^{\mathcal{S}} - m_i^{\mathcal{S}} m_j^{\mathcal{S}} \), where \( m_i^{\mathcal{S}} = \langle s_i \rangle^{\mathcal{S}} = \frac{1}{M} \sum_{\mu} s_i^\mu \) is the magnetization at site \( i \), and \( \chi_{ij}^{\mathcal{S}} = \langle s_i s_j \rangle^{\mathcal{S}} = \frac{1}{M} \sum_{\mu} s_i^{\mu} s_j^{\mu} \) is the total correlation between spins at sites \( i \) and \( j \). For the same of simplicity in notation, we drop the superscript $\mathcal{S}$ in the following calculations for the averages, i.e. $ m_i\equiv  m_i^{\mathcal{S}}$ and $\chi_{ij}\equiv \chi_{ij}^{\mathcal{S}}$.

The SM method provides approximations for the pairwise interaction matrix elements \( \{ J_{ij}^* \}_{i,j=1}^N \) and the external fields \( \{ h_i^* \}_{i=1}^N \) through the following relations (\( L_i \equiv 1 - m_i^2 \) and \( K_{ij} \equiv \frac{c_{ij}}{L_i L_j} \))~\cite{Sessak2009Small}:

\begin{widetext}
\begin{equation}\label{eq:sm4}
\begin{aligned}
J_{ij}^* (\{c_{kl}\}, \{m_i\}, \beta) &= \beta K_{ij} - 2 \beta^2 m_i m_j K_{ij}^2-\beta^2\sum_{\substack{k}} K_{jk} K_{ki} L_k + \frac{1}{3} \beta^3 K_{ij}^3 \left[ 1 + 3m_i^2 + 3m_j^2 + 9m_i^2 m_j^2 \right] \\
&\quad + \beta^3 \sum_{\substack{k \\ (\neq i, \neq j)}} K_{ij} \left( K_{jk}^2 L_j + K_{ki}^2 L_i \right) L_k 
+ \beta^3 \sum_{\substack{k,l \\ (k \neq i, l \neq j)}} K_{jk} K_{kl}K_{li} L_k L_l + O(\beta^4).
\end{aligned}
\end{equation}
\begin{equation}\label{eq:sm5}
\begin{aligned}
h_l^*(\{c_{ij}\}, \{m_i\}, \beta) &= \frac{1}{2} \ln \left( \frac{1 + m_l}{1 - m_l} \right) - \sum_j J_{lj}^* m_j + \beta^2 \sum_{j (\neq l)} K_{lj}^2 m_l L_j - \frac{2}{3} \beta^3 \left( 1 + 3m_l^2 \right) \sum_{j (\neq l)} K_{lj}^3 m_j L_j \\
&\quad - 2\beta^3 m_l \sum_{j < k} K_{lj} K_{jk} K_{kl} L_j L_k + O(\beta^4).
\end{aligned}
\end{equation}
\end{widetext}
These two equations are the building blocks of our SM method, employed in this study.
\subsection{\label{sec:sm4}Nodewise Logistic Regression}
Nodewise \( \ell_1 \)-regularized logistic regression, as proposed by Ravikumar et al.\cite{Ravikumar2010High}, was specifically designed for the inverse Ising problem in the absence of external fields. Under this condition, the probability distribution is given by:
\begin{equation}
    P_{\boldsymbol{J}}(\boldsymbol{s}) = \frac{\exp\left( \sum_{i,j=1}^{N} J_{ij} s_i s_j \right)}{Z(\boldsymbol{J})}
\label{eq:sm6}
\end{equation}
where \( Z(\boldsymbol{J}) \) is the partition function. Each spin \( s_g \) is treated as the dependent variable, while the remaining spins, \( s_{V \setminus \{g\}} = \{ s_i \}_{i \neq g} \), serve as the set of predictors. The pairwise coupling parameters for lattice site \( g \), denoted as \( \boldsymbol{J}_g = \{ J_{ig} \mid i \neq g \} \), are estimated by solving the following \( \ell_1 \)-regularized logistic regression optimization problem:
\begin{equation}\label{eq:sm7}
\hat{\boldsymbol{J}_g} = \arg \min_{\boldsymbol{J}_g \in \mathbb{R}} \left[ 
-\frac{1}{M} \sum_{i=1}^M \log P_{\boldsymbol{J}} \left( s_g^{(i)} \mid s_{V \setminus \{g\}}^{(i)} \right) 
+ \lambda \|\boldsymbol{J}_g\|_1 
\right]
\end{equation}
where \( M \) represents the number of independent and identically distributed samples, \(\| J_i \|_1 = \sum_{j \neq i} \| J_{ij} \| \) which avoids the interaction matrix from being dense, and \( \lambda \) is a user-defined penalty parameter~\cite{Ravikumar2010High}. In this research, the solutions to Eq.~\ref{eq:sm7} were computed for 100 different penalty values, chosen from a geometric sequence. The value of \( \lambda \) that resulted in the minimal extended Bayesian information criterion was selected as the optimal solution. For the implementation of this method, we utilized its MATLAB implementation as provided in the work by Brusco et al~\cite{brusco2023comparison}.

\begin{figure}[htbp]
    \centering
    \includegraphics[width=0.35\textwidth]{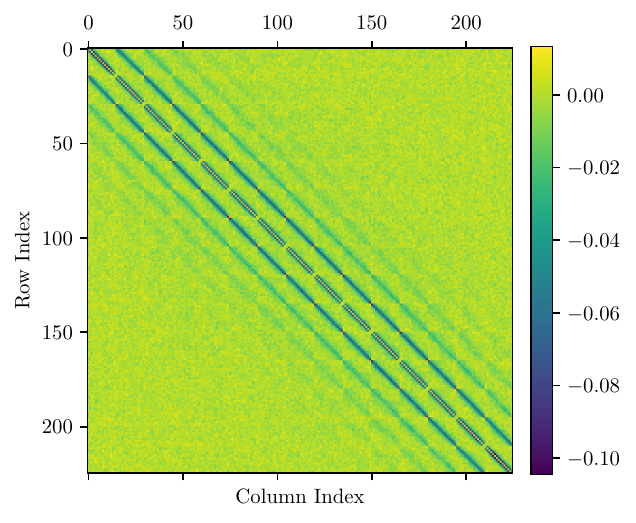}
    \caption{Interaction matrix estimated using the Sessak and Monasson method.}
    \label{fig:sm8}
\end{figure}

\begin{figure}[htbp]
    \centering
    \includegraphics[width=0.35\textwidth]{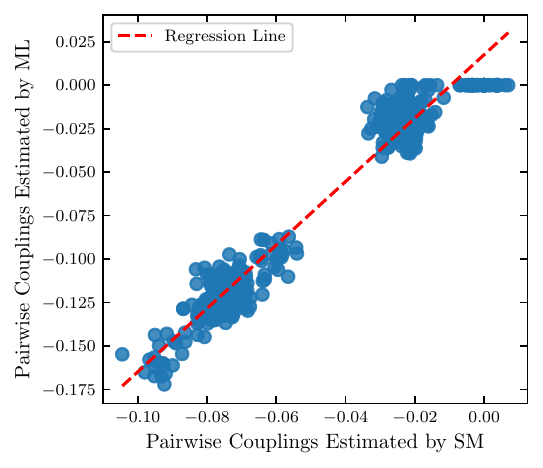}
    \caption{Pairwise couplings of two off-diagonal entities of the interaction matrix estimated by the ML approach, compared to those estimated by the SM method.}
    \label{fig:sm9}
\end{figure}

The estimated interaction matrix of a DSM with a lattice size of 256 and \(\zeta = 16\), calculated using the SM method, is shown in Fig. \ref{fig:sm8}. This indicates that the pairwise couplings weaken as they move away from the main diagonal.
We also compared the pairwise couplings estimated using the SM method and \( \ell_1 \)-regularized logistic regression optimization problem (machine learning, ML). As shown in Fig. \ref{fig:sm9}, the values obtained from the SM method tend to underestimate the couplings compared to the machine learning results, a point that has been highlighted in the research by Roudi et al~\cite{roudi2009statistical}.
It is important to note that we successfully mapped the BTW model onto the Ising model. For a lattice size of 15, the direct magnetization was found to be \( 0.18 \pm 0.04 \), while using the SM method with consideration of external fields, we obtained a value of \( 0.18 \pm 0.06 \).  The comparison of spin-spin correlations obtained directly from the BTW model and the SM method is presented in the SSC versus distance inset of Fig. \ref{fig:SSC} in the main text.
 
\end{document}